# Unraveling Nanoscale Magnetic Ordering in Fe$_3$O$_4$ Nanoparticle Assemblies via X-rays

**Karine Chesnel** [1,*], **Dalton Griner** [1], **Dallin Smith** [1], **Yanping Cai** [1], **Matea Trevino** [1], **Brittni Newbold** [1], **Tianhan Wang** [2], **Tianmin Liu** [2], **Emmanuelle Jal** [2], **Alex H. Reid** [2] and **Roger G. Harrison** [3]

[1] Department of Physics & Astronomy, Brigham Young University, Provo, UT 84602, USA; dalton.griner@gmail.com (D.G.); dallinsmith9@gmail.com (D.S.); kenzocyp@gmail.com (Y.C.); siameserulz@sbcglobal.net (M.T.); bnew94@gmail.com (B.N.)
[2] Stanford Institute for Materials and Energy Sciences, SLAC National Accelerator Laboratory, Menlo Park, CA 94025, USA; hahniel@gmail.com (T.W.); tmliu@slac.stanford.edu (T.L.); emmanuelle.jal@gmail.com (E.J.); alexhmr@slac.stanford.edu (A.H.R.)
[3] Department of Chemistry & Biochemistry, Brigham Young University, Provo, UT 84602, USA; rgharrison@chem.byu.edu
\* Correspondence: kchesnel@byu.edu; Tel.: +1-801-422-5687

**Abstract:** Understanding the correlations between magnetic nanoparticles is important for nanotechnologies, such as high-density magnetic recording and biomedical applications, where functionalized magnetic particles are used as contrast agents and for drug delivery. The ability to control the magnetic state of individual particles depends on the good knowledge of the magnetic correlations between particles when assembled. Inaccessible via standard magnetometry techniques, nanoscale magnetic ordering in self-assemblies of Fe$_3$O$_4$ nanoparticles is here unveiled via X-ray resonant magnetic scattering (XRMS). Measured throughout the magnetization process, the XRMS signal reveals size-dependent inter-particle magnetic correlations. Smaller (5 nm) particles show little magnetic correlations, even when packed close together, yielding to magnetic disorder in the absence of an external field, i.e., superparamagnetism. In contrast, larger (11 nm) particles tend to be more strongly correlated, yielding a mix of magnetic orders including ferromagnetic and anti-ferromagnetic orders. These magnetic correlations are present even when the particles are sparsely distributed.

## 1. Introduction

Magnetic nanoparticles have become increasingly useful in nanotechnologies [1–3]. Their strong, flippable magnetization can be used in magnetic recording technologies and their small size enables higher storage densities [4]. Additionally, magnetic nanoparticles have numerous applications in biomedicine [5,6]. In magnetic resonance imaging (MRI), they are used as contrast enhancement agents [7]. In drug and gene delivery, high-moment magnetic nanoparticles with multifunctional coatings are used as carriers to deliver specific molecules to target sites [8]. The magneto-thermic properties of magnetic nanoparticles are also exploited for therapeutic hyperthermia to treat cancer and tumors [9].

Magnetite (Fe$_3$O$_4$) nanoparticles are particularly appealing because of their biocompatibility. They are non-toxic and have a long lifetime in the bloodstream [10–12]. Fe$_3$O$_4$ carries a strong magnetization, supported by the spins of two ions, Fe$^{2+}$ and Fe$^{3+}$, distributed ferrimagnetically throughout a spinel crystallographic structure [13,14]. Nano-sized $Fe_3O_4$ particles exhibit a strong bulk-like magnetic moment throughout the core of the particle with limited oxidation effects at the surface [15–17]. When their size is below about 100 nm, $Fe_3O_4$ particles are usually magnetically monodomain, thus forming a single giant magnetic moment, or "nanospin" [18,19].

A collection of monodomain $Fe_3O_4$ nanoparticles often exhibits superparamagnetism (SPM), where, in the absence of an external magnetic field, individual nanospins tend to be randomly oriented [3,20]. Under the application of an external field, the nanospins tend to align in the direction of the field. The field magnitude necessary to align all the particles depends on particle size and on temperature [21,22]. A number of magnetometry and spectroscopy measurements conducted on $Fe_3O_4$ nanoparticles of various sizes have helped determine physical parameters, such as the magnetic anisotropy, the coercive field, and the blocking temperature $T_B$, marking the transition between the SPM state (above $T_B$) and the magnetically blocked state (below $T_B$) at a given timescale [23–25]. Most of these investigations have focused on the net macroscopic magnetization of the material.

Although crucial for nano-technological applications, information about the magnetic ordering of the individual nanospins at the nanoscale is often lacking because of the difficulty to access it experimentally. Standard magnetic force microscopy (MFM) does not currently allow resolving magnetic features smaller than ≈25 nm. Recently, the development of transmission electron microscopy (TEM) tuned to Fresnel Lorentz microscopy (FLM) mode and electron holography (EL) mode has allowed for the pioneer imaging of the magnetic state of particles as small as 15 nm Co particles [26], 13 nm $Fe_3O_4$ particles [27], and 7 nm $Fe_3O_4$ particles [28]. These exciting real-space imaging techniques however require very thin materials (a monolayer of particles or less) and are essentially sensitive to the magnetization orientation in the plane of the layer.

Exploiting the sensitivity of the neutron to magnetic moments, small-angle neutron scattering (SANS) has been very helpful to probe nanoscale magnetic correlation lengths in large collections of nanoparticles [29–31]. However, in order to obtain a sufficient signal-to-noise ratio, SANS generally necessitates large amounts of material, and using SANS to probe thin layers of nanoparticles is challenging. Grazing incidence polarized neutron reflectivity (PNR), on the other hand, has recently provided data on thin monolayers of iron oxide nanoparticles, of 25 nm [32] and 18 nm [33]. While the PNR technique allows to extract nuclear and magnetic in-depth profiles, it does however not provide information about the lateral magnetic structure in the plane of the layer.

With wavelengths in the nanometer range, soft X-rays produced by synchrotron radiation are brilliant enough to produce, when tuned to absorption resonances, an X-ray resonant magnetic scattering (XRMS) signal from thin nanometric magnetic films and thus provide information about lateral magnetic correlations in nanoparticle assemblies not accessible by other techniques. Complementary to electron holography techniques, and to neutron scattering techniques, XRMS allows access to the magnetization component, not only in-plane, but also out-of-plane. Furthermore, the resonant nature of XRMS and its dependence on light polarization allows the simultaneous collection of charge and magnetic scattering signals providing information on both the charge correlations and the magnetic correlations in the nanoparticle assemblies [34].

Our XRMS study conducted on self-assemblies of $Fe_3O_4$ nanoparticles yields information about the lateral nanoscale magnetic ordering between the particles. By tuning the energy of the X-rays to magnetic resonances and exploiting the light polarization, we extracted a pure magnetic scattering signal. The shape of this magnetic signal reveals nanoscale magnetic orders such as "ferromagnetic" (parallel alignment) and "antiferromagnetic" (antiparallel alignment) orders, as well as magnetic randomness, which is a lack of ordering. We measured the magnetic scattering signal at several field values and found an interesting evolution throughout the magnetization process. Additionally, we have observed that the magnetic ordering behavior depends on particle size.

## 2. Results

*2.1. Structural Properties of the Nanoparticle Assemblies*

We present here results for two different particle sizes: 5 nm (labelled M5) and 11 nm (labelled M11). Figure 1A,B shows transmission electron microscopy (TEM) images of these two particle sizes where a drop of their respective solutions was deposited onto silicon nitride $Si_3N_4$ membranes. Consistent with XRD measurements [22], the statistical analysis of the TEM images shows that a better size control was achieved for the smaller particles M5, with an average size $D \approx 5.5 \pm 0.8$ nm. Wider size variations were observed for the larger particles M11, with an average size $D \approx 11.0 \pm 4.6$ nm. The nanoparticles tend to self-assemble in hexagonal arrangements. In the case of M5, the narrower size distribution allowed a closer packing and more uniform coverage, whereas in the case of M11, the wider size distribution led to a looser packing with the formation of islands, separated by empty regions.

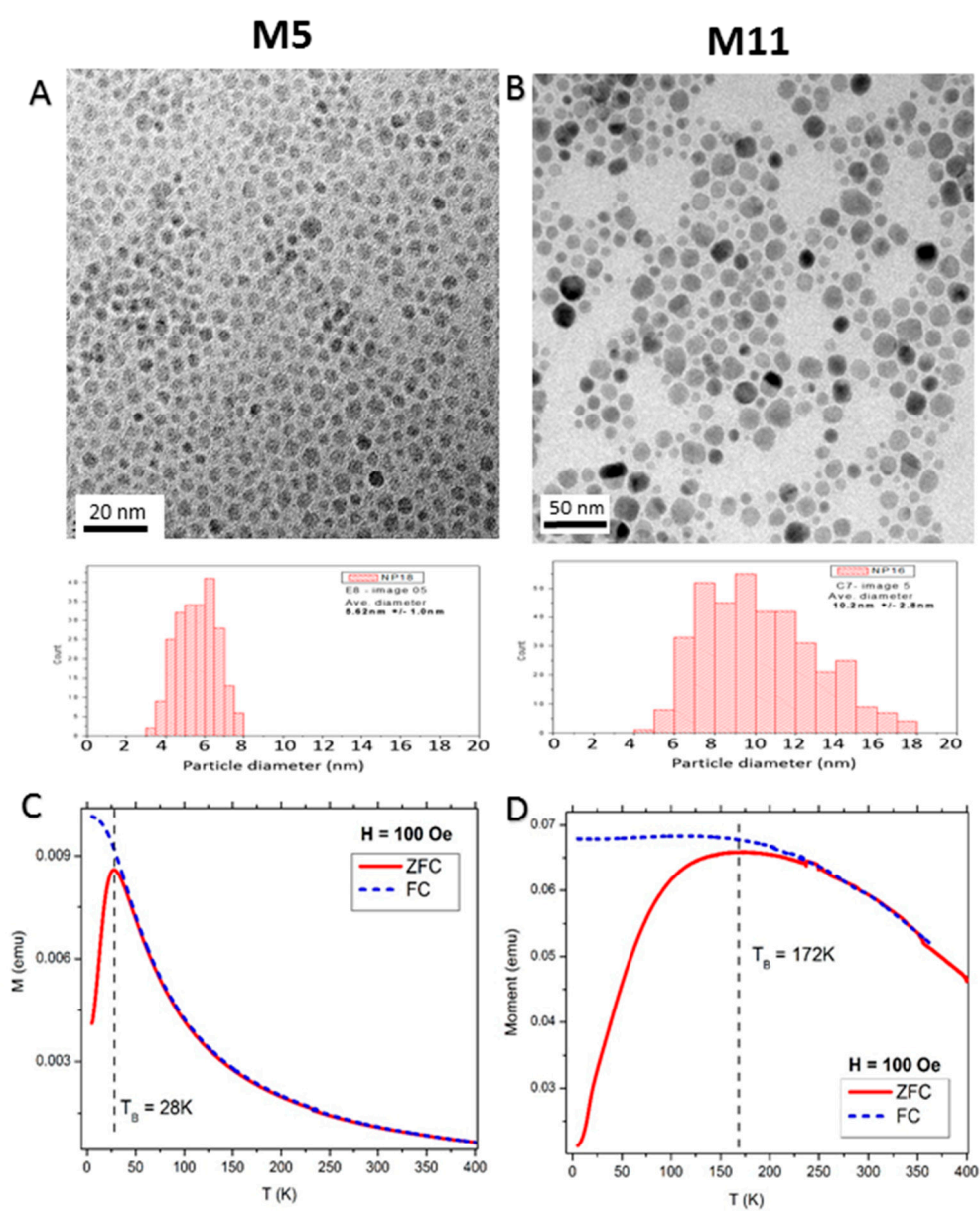

**Figure 1.** Views of the $Fe_3O_4$ nanoparticle assemblies and magnetometry data for M5 and M11. (**A**,**B**) Transmission electron microscopy (TEM) images of M5 (**A**) and M11 (**B**), with associated size distribution histograms below each image. (**C**,**D**) Magnetization vs temperature curves measured in zero-field-cooling (ZFC, red curves) and field cooling (FC, blue curves) conditions under a field of 100 Oe, for M5 (**C**) and M11 (**D**).

*2.2. Macroscopic Magnetic Properties*

Magnetometry measurements were carried out on both particle sizes. In Figure 1C,D, field-cooling (FC) and zero-field-cooling (ZFC) curves measured at $H$ = 100 Oe show a drastic dependence on particle size. When a field of 100 Oe was applied, M5 evolved from the SPM state to the magnetically blocked state at a blocking temperature $T_B$ = 28 K, while for M11, this transition occurred at a much higher temperature around $T_B$ = 172 K. This drastic difference in $T_B$ indicates that particle size impacts the strength of the magnetic interactions between the nanoparticles. Inter-particle magnetic interactions appear stronger for M11 than for M5, as, in case of M11, the temperature needs to be further raised for thermal fluctuations to overcome magnetic interactions. Consequently, the local magnetic ordering between the nanoparticles is likely to depend on the particle size too.

*2.3. X-ray Scattering Patterns: Nanoscale Charge Correlations*

The self-assemblies of $Fe_3O_4$ nanoparticles were probed via X-ray resonant magnetic scattering (XRMS) using synchrotron radiation [35–38]. For this purpose, the energy of the X-rays was carefully tuned to the $L_3$ resonance edge of Fe, at around 707 eV, in order to enhance the magneto-optical contrast and access magnetic information in the material [39,40]. The particles were deposited onto silicon nitride $Si_3N_4$ membranes to allow the collection of scattering patterns in transmission geometry as shown in Figure 2A. During the XRMS measurement, the 20 × 30 μm membrane windows were entirely illuminated by the X-ray beam, whose transverse size was ≈200 × 80 μm, thus covering several million particles, including regions of various packing orientations. Consequently, like in powder diffraction, the resulting scattering signal is isotropic, with no preferred orientation. When measured on the detector, the scattering signal has the shape of a ring (intersection between the scattering cone and the detector plane) as illustrated in Figure 2. The detector, a charge-coupled-device (CCD) camera with 2048 × 2048 pixels of 13.5 μm in size, was placed at 105 mm downstream of the sample. To prevent intense direct X-rays from damaging the CCD detector, a beamstop was used, in the shape of a disk mounted on a rod. The shadow of the beamstop is visible at the center of the images collected on M11. For measurements on M5, the detector was shifted sideways and a portion of the beamstop is visible on the left side of the images.

Scattering patterns produced by M5 and M11 (Figure 2B,C) exhibit rings of different sizes. The radius of the ring is related to the average inter-particle distance in an inverse relationship. For the bigger particles M11, the ring was entirely captured by the detector, whereas for the smaller particles M5, the ring radius was much larger, and as a result, only a portion of the ring was captured. To capture as much of the scattering ring as possible, the detector was shifted sideways. We integrated the 2-D scattering patterns angularly on rings of constant radius, i.e., scattering vector $q$, so as to produce 1-D scattering profiles $I(q)$ (Figure 2D,E). Prior to this integrating, we cleaned the initial 2-D images by zero-ing the intensity in all the pixels situated in the shadow of the blocker, so no artificial signal from the blocked region was included in the 1-D scattering profiles. The resulting profiles for M5 (Figure 2D) and for M11 (Figure 2E) both exhibit a peak. The location of the peak $q^*$ indicates the average inter-particle distance $p = 2\pi/q^*$ which is the average distance from the center of one particle to the center of the nearest neighboring particle in the given particle assembly. The distance $p$ therefore includes the particle diameter $D$ and the spacing between particles $L$ ($p = D + L$). The width $\Delta q$ of the peak indicates the correlation length $\lambda_c = 2\pi/\Delta q$, i.e., the distance over which the particles' positions are correlated in any given direction. For M11, a large diffuse scattering background existed below the peak. For the sake of evaluating the position $q^*$ and the width $\Delta q$ of the peak accurately, the diffuse background was previously removed. The XRMS peak location $q^*$ yielded $p ≈ 6.25$ nm for M5 and $p ≈ 19.00$ nm for M11. For M5, $\lambda_c$ = 49 nm, covering about eight particles, whereas for M11, $\lambda_c$ = 66 nm, covering about 3.5 particles. In other words, the peak's width $\Delta q$ provides an uncertainty on $p$. In the case of M5, $p = 6.25 ± 0.86$ nm, while in the case of M11, $p = 19.0 ± 5.45$ nm. This agrees with earlier TEM measurements (Figure 1A,B), from which the diameter $D$ was extracted. The M5 particles, for which the average diameter is $D \approx 5.5$ nm, are rather close-packed, with an average separation of only $L = p − D \approx 0.75$ nm, whereas the M11 particles, for which the average diameter is $D \approx 11$ nm, are more loosely packed with an average separation $L = p − D \approx 8$ nm.

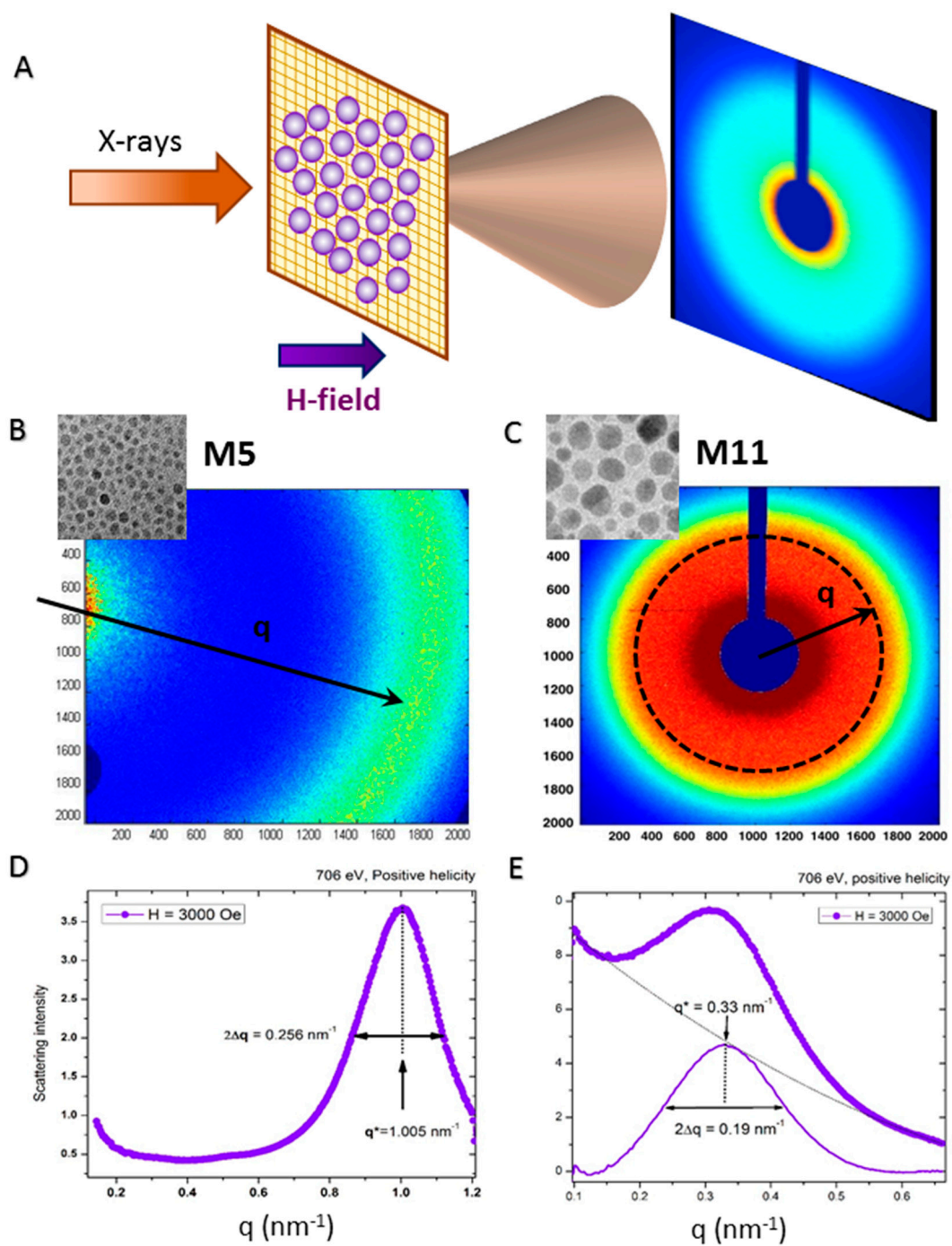

**Figure 2.** X-ray resonant magnetic scattering (XRMS) measurements on the $Fe_3O_4$ nanoparticle assemblies carried at the $L_3$ resonant edge of Fe (707 eV). (**A**) Layout of the XRMS experiment showing nanoparticle assemblies placed in transmission geometry and scattering signal collected downstream with a CCD detector of 2048 × 2048 pixels of 13.5 μm placed at 105 mm away from the sample. (**B**) XRMS pattern for M5. (**C**) XRMS pattern for M11. The dark blue region at the center is the shadow of a beamstop preventing intense direct X-rays from damaging the detector. (**D**) Scattering intensity profile $I(q)$ for M5 obtained by angularly integrating pattern B. (**E**) Scattering intensity profile $I(q)$ for M11 obtained by angularly integrating pattern C. The graph also includes a modified signal corrected from the background for the purpose of peak positioning only.

*2.4. Extracted Magnetic Scattering: Nanoscale Magnetic Correlations*

The XRMS scattering signal observed here is mainly produced by charge scattering, i.e., the electron density distribution in the material. However, because the X-ray energy is tuned to the Fe $L_3$ resonance, a portion of the scattering signal is magnetic and contains information about magnetic ordering [41].

According to the magnetic scattering theory [39,40], the XRMS intensity $I(q)$ depends on the polarization of the X-ray light. As shown in Appendix A, an expression for $I(q)$ may be derived from scattering factors $f$, which include a charge component $f_c$ and a magnetic component $f_m$. When measured in circularly polarized light, the scattering amplitude $A(q)$ can be written as a sum of two terms:

$$A_{\pm}(q) = f_c s_c \pm f_m s_m \qquad (1)$$

where the $\pm$ sign refers to the two opposite helicities of the light polarization, and the coefficients $s_c$ and $s_m$ contain information about the spatial distribution of the charge and the magnetic densities, respectively. The resulting scattering intensity $I_{\pm}(q) = |A_{\pm}(q)|^2$ will also be composed of a combination of charge and magnetic terms.

To extract information about the magnetic distribution, we compare XRMS patterns collected in opposite light helicities and their associated intensities $I_+$ and $I_-$. For each comparison, we compute two ratios, a dichroic ratio $R_D$ and a magnetic ratio $R_M$, defined as follows:

$$R_D = \frac{I_+ - I_-}{I_+ + I_-} \qquad \text{and} \qquad R_M = \frac{I_+ - I_-}{\sqrt{(I_+ + I_-)}} \qquad (2)$$

As shown in the Appendix A, $R_D$ is roughly proportional to the magnitude of the ratio $s_m/s_c$, whereas $R_M$ is roughly proportional to the magnitude of $s_m$ only. Like $I_+$ and $I_-$, the ratios $R_D$ and $R_M$ depend on $q$. Plotting $R_M(q)$ provides an image of the magnetic structure factor $s_m(q)$, revealing any magnetic ordering. Note that the intensities $I_+$ and $I_-$ used for the calculation of $R_D$ and $R_M$, are the raw intensities normalized by the intensity $I_0$ of the incoming light illuminating the material. No prior background removal (such as the one shown in Figure 2E) was applied. Indeed, it is expected that the diffuse background includes precious charge and magnetic scattering information. The purpose for calculating $R_D$ and $R_M$ is precisely to extract magnetic information in the diffuse background signal for the entire range of $q$ accessible in the measured data, especially at $q < q^*$.

Figure 3A shows $I_+(q)$ and $I_-(q)$ profiles measured at 708 eV (where the magnetic contrast is enhanced as explained in next paragraph) for M11 under a magnetic field $H$ = 3000 Oe. The associated ratio $R_D(q)$ in Figure 3B shows an average value of 26%, indicating a significant magnetic component in the XRMS signal. $R_D(q)$ is essentially constant with respect to $q$, suggesting that at that field value, the magnetic and structural orders are similar. Unlike $R_D(q)$, the ratio $R_M(q)$ in Figure 3C significantly varies with $q$. The shape of $R_M(q)$ mimics that of $I_{\pm}(q)$ with a peak at the same location $q^*$ (the value of $q^* \approx 0.38$ nm$^{-1}$ is slightly higher than for Figure 2E because the M11 particles measured in Figure 3 were more densely packed, with an average inter-particle distance $p \approx$ 16.5 nm). The matching in peak position between $R_M(q)$ and $I_{\pm}(q)$ suggests that at $H$ = 3000 Oe, the assembly of particles exhibited a magnetic order, for which the magnetic period $p_m$ is the same as the structural period or inter-particle distance $p$. Such magnetic order corresponds to a parallel alignment of the nanospins in the direction of the external field $H$. In the rest of this manuscript, we label this order as "ferromagnetic", assuming a parallel, or ferromagnetic alignment of the nanospins at the nanoscale. Indeed, at $H$ = 3000 Oe, the magnetometry data for M11 shown in Figure 4 suggests that the material is nearly saturated, so nearly all the particles are aligned with the external field.

To verify the magnetic nature of $R_M(q)$, we looked at its dependence with the X-ray energy. The X-ray absorption spectroscopy (XAS) signal (Figure 3D) shows an absorption peak at the Fe $L_3$ edge at ≈707 eV and $L_2$ edge at ≈720 eV. The associated X-ray magnetic circular dichroic (XMCD) signal exhibits three peaks of alternating sign at energies $E_1$ = 706 eV, $E_2$ = 707 eV, and $E_3$ = 708 eV. This XMCD signal is caused by the magnetic moments carried by the Fe atoms in the material. The three distinct peaks are the signature of the chemical compound $Fe_3O_4$ where the $Fe^{2+}$ and $Fe^{3+}$ ions occupy three different crystallographic sites [42]. To optimize the magneto-optical contrast in XRMS, the energy of the X-rays was finely tuned to either $E_1$, $E_2$, or $E_3$. We measured $R_M(q)$ at these three energies (Figure 3E). Strikingly, the sign alternation seen in the XMCD signal (Figure 3D) was also observed in the $R_M(q)$ profiles (Figure 3E): $R_M(q)$ was all positive at $E_1$ and $E_3$ but all negative and its shape was flipped at $E_2$. This result confirms that $R_M(q)$ is of magnetic nature.

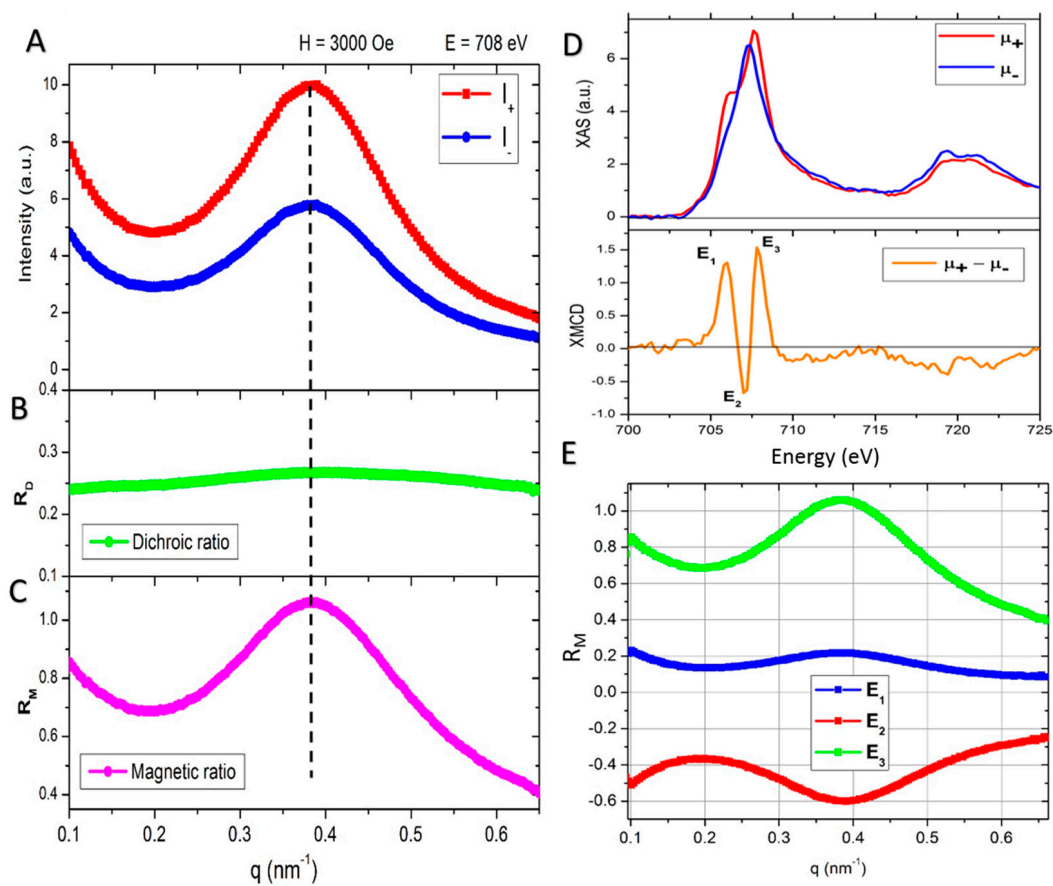

**Figure 3.** Extraction of magnetic scattering profiles and dependency with the X-ray energy for M11. (**A**) Scattering intensity profiles $I_+(q)$ and $I_-(q)$ collected in opposite helicities. (**B**) Associated dichroic ratio $R_D(q)$. (**C**) Associated magnetic ratio $R_M(q)$. (**D**) X-ray absorption spectroscopy (XAS) spectra collected in opposite helicities, with associated X-ray magnetic circular dichroic (XMCD) signal. (**E**) Magnetic ratio $R_M(q)$ profiles measured at the energies $E_1$, $E_2$, and $E_3$ of the XMCD peaks marked in (D).

*2.5. Dependence on External Magnetic Field*

We measured $R_M(q)$ at $E_1 = 706$ eV for different field values, from $H = 0$ to 3000 Oe, throughout the magnetization process at room temperature (300 K). Results for M5 are shown in Figure 4A–C. The magnetization curve in Figure 4A shows a shape typical of SPM behavior. The normalized magnetization $M/M_s$ was about 60% at $H = 3000$ Oe and gradually decreased as $H$ decreased to reach about zero at $H = 0$. The red dots on the magnetization curve indicate points where XRMS was collected. The associated $R_M(q)$ profiles shown in Figure 4B all exhibit a peak at $q^* \approx 1$ nm$^{-1}$, where the charge scattering peak in Figure 2D is also located. At $H = 3000$ Oe, where the peak is the strongest, a majority of the nanospins are aligned with the external magnetic field $H$ so the magnetic density distribution mostly matches the charge density distribution. The average magnetic period, $p_m = 6.36 \pm 0.89$ nm, indicates a ferromagnetic (FM) order extending over about seven particles. When $H$ is decreased to lower values, the magnitude of $R_M(q)$ also decreases. In a fashion roughly proportional to the magnetization $M$, the magnitude of peak in $R_M(q)$ progressively decreases and eventually disappears when $H \approx 0$. In Figure 4C, a comparative plot of the $R_M(q)$ profiles where the $q^*$ peak has been normalized to 1 shows that throughout most of the range of field values $H$, the shape of $R_M(q)$ remains essentially the same. The main peak at $q^*$ remains the prevalent feature, suggesting that the FM order is the sole order existing in the material while a significant field $H$ is applied. Only at small field values, $H$ = 400 G and below, a secondary peak at $q^*/2 \approx 0.5$ nm$^{-1}$ arises, which suggests the appearance of some antiferromagnetic (AFM) arrangement between particles, for which the magnetic period $p_m$ = 2 $p$. However, the height of the AFM peak at $q^*/2$ is relatively small. At $H$ = 200 G, the height of the AFM peak (at $q^*/2$) is less than half (about 2/5) of the FM peak's height (at $q^*$), which itself has significantly reduced compared to the peak height at other field values (Figure 4B). Indeed, at $H$ = 200 G, the net magnetization $M/M_s$ of the material (Figure 4A) is only 5%, so the extent of the FM order is about 5% and, consequently, the extent of the AFM order was less than 2% of the nanoparticle assembly. This suggests that while there might still be islands that are either FM or AFM ordered, most of the nanospins are randomly oriented when the field $H$ is decreased down to zero. This confirms the SPM behavior of the nanoparticle assembly M5.

A similar study was conducted on M11, shown in Figure 4D–F. The magnetization curve in Figure 4D indicates that at $H$ = 3000 Oe, the magnetization is nearly complete, with $M/M_s \approx 90\%$. The associated magnetic profiles $R_M(q)$ in Figure 4D show a different dependence on $H$ than for M5. At $H$ = 3000 Oe, $R_M(q)$ exhibits a peak, located at $q^* \approx 0.33$ nm$^{-1}$, where the charge scattering peak in Figure 2E is also located, thus revealing an FM order. However, when $H$ is decreased, the shape of $R_M(q)$ significantly changes. This change is highlighted in Figure 4F, showing a comparative plot of the $R_M(q)$ profiles where the $q^*$ peak has been normalized to 1. The peak at $q^*$ progressively disappears and transforms into a plateau, spreading over lower $q$ values. The occurrence of a magnetic signal in the low $q < q^*$ region suggests the progressive loss of the FM order and the occurrence of a different magnetic ordering where $p_m > p$. An example of such ordering is the AFM order, producing a signal around $q^*/2$ in $R_M(q)$. The spreading out in the signal suggests a mix of different magnetic orders for the assembly of M11 particles. Even though no narrow AFM peak is visible at $q^*/2$, the relative magnitude of the low $q$ signal for M11 is much stronger than the relative magnitude of the AFM peak in M5. At $H$ = 100 G, the net magnetization $M/M_s$ in the material (Figure 4D) is still about 19%, but the $R_M(q)$ signal is not any more concentrated on the FM peak and has significantly spread over lower $q$ values. The magnitude of the combined lower $q$ signal is about the same as the magnitude of the FM peak. This indicates that while about 19% of the nanospins assembly is FM ordered, another significant portion (in the range of 20% or higher) of the assembly is exhibiting other magnetic orders, including AFM order. At $H$ = 0, the combined regions of AFM and other orders leads to roughly no net magnetization ($M \approx 0$). In summary, the nanospins in M11 exhibit significant magnetic correlations that compete with the SPM randomness, but still allow for the net magnetization $M$ to reach about zero when the external field was brought back to $H$ = 0.

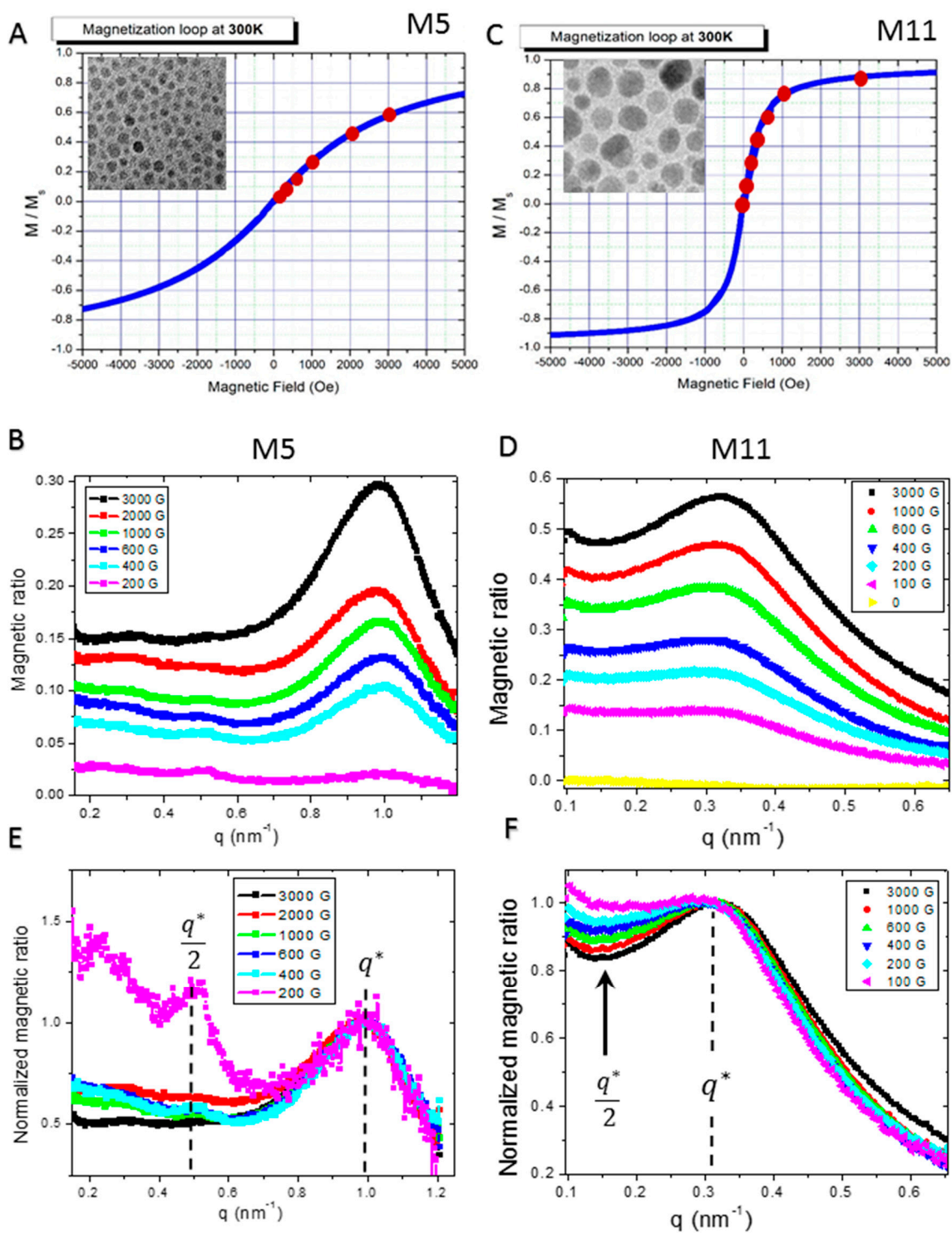

**Figure 4.** Magnetic profiles $R_M(q)$ measured at $E_1$ = 706 eV, at various magnetic fields throughout the magnetization reversal. (**A**) Magnetization curve measured at 300 K on M5. (**B**) $R_M(q)$ profiles for M5 measured at various field values $H$, indicated by the red dots on (**A**). (**C**) Comparative plot of the $R_M(q)$ profiles for M5 after normalizing the main peak at $q^* \approx 1$ nm$^{-1}$ to 1. Note that, through this normalization process, the magnitude of the signal for $H$ = 200 G was artificially amplified. The secondary peak appearing at $q^*/2 \approx 0.5$ nm$^{-1}$ is actually very small in magnitude (less than 2% of the magnitude of the peak at $q^*$ for $H$ = 3000 G) (**D**) Magnetization curve measured at 300 K on M11. (**E**) $R_M(q)$ profiles for M11 measured at various field values $H$, indicated by the red dots on (**C**). (**F**) Comparative plot of the $R_M(q)$ profiles for M11 after normalizing the main peak at $q^* \approx 0.33$ nm$^{-1}$ to 1.

## 3. Discussion

This XRMS study suggests that, when present, the magnetic ordering in $Fe_3O_4$ nanoparticle assemblies drastically depends on particle size. For the smaller particles M5, the only significant magnetic order visible in the XRMS signal is the FM order, where nanospins align with each other and for which the magnetic period $p_m$ matches the inter-particle distance $p$. This FM order only occurs in the presence of an external magnetic field $H$. The extent of the ferromagnetic order throughout the particle assembly is set by the magnitude of $H$. At high, nearly saturating field $H$, nearly all the nanospins are aligned with the field, and the ferromagnetic order covers nearly 100% of the assembly. When the field $H$ was decreased, some nanospins start to reverse and magnetically fluctuate, so the ferromagnetic coverage $c$ decreases in a manner roughly proportional to the net magnetization ($c \propto M$). At $H$ = 3000 Oe, the net magnetization $M$ equates to 60% of the magnetization at saturation $M_s$. Consequently, the extent of the FM coverage $c$ is about 60%. When the field approaches zero, the FM order progressively disappears ($c \approx 0$). At very low magnetization levels ($M/M_s < 5\%$), a small portion of the nanospin assembly (likely located in sparse islands) exhibits AFM order but the majority of the nanospins are randomly aligned, leading to zero net magnetization ($M \approx 0$). In the absence of an external field, magnetic randomness prevails. This confirms that at room temperature, well above its blocking temperature ($T_B$ = 28 K) M5 is SPM. Even though the particles are tightly close-packed, the individual nanospins show little correlation, and randomly fluctuate in the absence of a magnetic field. Only a strong external magnetic field could cause the nanospins to align. The ferromagnetic order is induced purely by the external field and the order completely disappears when the field is absent, leaving mostly disorder.

The bigger particles, M11, on the other hand, exhibit a variety of magnetic orderings. At $H$ = 3000 Oe, the net magnetization $M$ is 90% of the magnetization at saturation $M_s$, suggesting a FM coverage of $c \approx 90\%$. When $H$ is lowered, $M$ gradually decreases, and so does the FM coverage $c$. However, unlike for M5, when the field $H$ approaches zero, the magnetic correlations do not completely disappear and the nanospins do not all randomly fluctuate. The appearance of a strong XRMS signal at different length scales suggests magnetic orderings with a magnetic period $p_m$ larger than the inter-particle distance $p$. This suggests the presence of magnetic correlations between neighboring particles through dipolar couplings that compete with thermal magnetic fluctuations. The nanospins tend to partially align, even in the absence of a field. When $H \approx 0$, a portion of the nanospins order preferentially in an AFM arrangement. With the alternation of spin up and down, the AFM order leads to zero net magnetization like randomness does. The AFM ordering allows the material to achieve zero net magnetization even in the presence of magnetic couplings between particles. In reality, given the loose arrangement of particles in M11 assembly, there is a mix of regions with AFM, FM, and other orders, as well as regions of magnetic randomness, all combining to produce a zero net magnetization ($M \approx 0$).

Despite a loose arrangement, it appears that the bigger M11 particles somewhat correlate magnetically, whereas even when tightly close-packed, the smaller M5 particles show little magnetic correlation and mostly show a SPM behavior when above $T_B$. This drastic difference may be due to the drastic change in the strength $M_n$ of the magnetization carried by individual nanospins when the size of the particle changes. $M_n$ can be evaluated by multiplying the estimated number of $Fe_3O_4$ molecules included in a particle by the magnetic moment per $Fe_3O_4$ formula unit (in $\mu_B$/f.u.), which can be measured using XMCD [17,43]. Our estimation of the number of $Fe_3O_4$ molecules is based on crystallographic data for $Fe_3O_4$, which has a spinel structure whose unit cell is 8.4 Å in size corresponding to a unit volume of 592 Å$^3$ and including four $Fe_3O_4$ molecules [44]. When the particle is about 5 nm in diameter, it includes about 588 $Fe_3O_4$ molecules. With a reported averaged moment of $\approx 2.5$ $\mu_B$/f.u., its estimated nanospin magnetization is $M_n \approx 1400$ $\mu_B$. On the other hand, if the particle is about 11 nm in diameter, it includes about 4725 $Fe_3O_4$ molecules. With a reported average moment of $\approx 3$ $\mu_B$/f.u., the estimated nanospin strength is $M_n \approx 14{,}000$ $\mu_B$, which is 10 times higher than for 5 nm particles. Our XRMS measurements suggest that this change in magnetization strength $M_n$ significantly impacts the ability for neighboring particles to correlate magnetically. In the absence of an external magnetic field, assemblies of bigger magnetic nanoparticles may show a variety of magnetic orderings competing with random thermal magnetic fluctuations, while assemblies of smaller particles may exhibit a pure SPM behavior.

These observations open the door to interesting questions, including dependence on particle size, concentration, and temperature. Next possible investigations include finer size-dependence studies and the effect of the nanoparticles concentration in the assemblies. For a given size and concentration, it would be useful to study the evolution of the magnetic ordering with temperature as the particles are cooled down below $T_B$. Also, a model of the nanospin assemblies would allow fitting the XRMS scattering signal and quantifying the coverage for the various magnetic orders. Combined with magnetometry and SANS data, the XRMS technique provides unique experimental information, useful for the development of theoretical models [45,46] and further understanding the behavior of magnetic nanoparticle assemblies at the nanoscale.

## 4. Materials and Methods

Our experimental work included the fabrication of $Fe_3O_4$ nanoparticles of different sizes on which we conducted the structural, magnetic, and X-ray scattering measurements described below.

The $Fe_3O_4$ nanoparticles were fabricated following organic routes [47,48] based on a thermal decomposition of an Fe oleate precursor, as described in our prior publication [22]. These methods provided some size control. The size of the nanoparticles was tuned mainly by adjusting the temperature and the length of time at which the solution was left refluxing under nitrogen before being cooled down and nanoparticles were precipitated by centrifugation in ethanol.

Once precipitated in a powder form, magnetization and field-cooling measurements were carried out via magnetometry. From the solution form, particles were deposited onto $Si_3N_4$ membranes and imaged via transmission electron microscopy (TEM). The deposited $Fe_3O_4$ particles were finally studied via X-ray resonant magnetic scattering (XRMS) using synchrotron radiation.

The magnetometry measurements were carried out with a Vibrating Sample Magnetometer (VSM) on a Physical Property Measurements System (PPMS) from Quantum Design (San Diego, CA, USA). The PPMS instrument included a superconducting magnet allowing measurements up to 9 T (90,000 Oe) and sample cooling down to a few kelvins. Our field-cooling (FC) and zero-field-cooling (ZFC) measurements were carried out by cooling samples down to 5 K under various field values, ranging from 0 to 500 Oe.

The imaging of the nanoparticles was done via TEM. The TEM instrument, a TF30 manufactured by FEI Co. (Portland, OR, USA), used a 300 keV electron beam. For the TEM imaging, the nanoparticles of various sizes and concentrations were deposited on silicon nitride ($Si_3N_4$) membranes that were 50 to 100 nm thick. The membranes had a 20 × 30 μm window in their center to let the electrons and the X-ray beams pass through.

The X-ray resonant magnetic scattering (XRMS) measurements were carried out at beamline 13.3 at the Stanford Synchrotron Radiation Light Source (SSRL) at the Stanford Linear Accelerator (SLAC) (Stanford, CA, USA). The energy of the X-rays was finely tuned to the $L_3$ edge of Fe at around 707 eV. The X-ray light, produced by an elliptical undulator (EPU), was polarized either circularly or linearly with a degree of polarization close to 98%. For dichroism studies, the helicity of the light was regularly switched from left to right by changing the EPU settings. The transverse size of the beam at the sample location was about 200 μm × 80 μm, thus largely covering the sample membranes when optimally aligned. The samples were mounted on a cryogenic sample holder to allow sample cooling with liquid helium down to about 15 K. The sample holder was inserted in custom-designed pole pieces of an electromagnet allowing XRMS measurements with an in situ magnetic field up to 3000 Oe. The scattering patterns were collected on a 2-D detector, a CCD camera with 2048 × 2048 pixels from Princeton Instruments (Trenton, NJ, USA) placed at 105 mm downstream of the sample.

Scattering images were analyzed and modeled at Brigham Young University. An azimuthal integration was applied to the 2-D images so to produce 1-D scattering profiles. Profiles were all individually normalized by a photocurrent signal ($I_0$) detected on the incident synchrotron beam, upstream of the samples at the time of the XRMS measurement. For each sample, at each in situ magnetic field magnitude, normalized profiles were carefully compared at different polarizations and X-ray energies.

## 5. Conclusions

In conclusion, our X-ray magnetic scattering study of $Fe_3O_4$ nanoparticle self-assemblies revealed various magnetic orderings at the nanoscale and their drastic dependence on particle size. We found that, even when tightly close-packed, the 5 nm $Fe_3O_4$ particles show pure superparamagnetic behavior as they do not correlate magnetically, leading to random magnetic fluctuations in the absence of external field. We found, on the other hand, that 11 nm particles tend to correlate magnetically through dipolar magnetic couplings when they are self-assembled, even if their arrangement is loose. If the external magnetic field is not strong enough to align the particles ferromagnetically, the dipolar magnetic couplings between particles lead to a variety of magnetic orders, with a preference for an antiferromagnetic order, competing with the random magnetic thermal fluctuations. These results open the door to further investigations of nanoscale magnetic ordering in nanostructured materials and their dependence on the temperature and external magnetic field. Furthermore, more information about the spatiotemporal behavior of the nanoparticles, such as their dynamics of fluctuation, can be obtained via the use of coherent X-rays for magnetic scattering [38] and speckle correlation techniques [49,50]. Coherent X-ray magnetic scattering also provides promising ways to visualize the magnetic structures in real space via coherent X-ray diffractive imaging [51]. In an era of fast developing nanotechnologies, these specialized techniques can help better understand the magnetic behavior of nanostructured materials at the nanoscale. This information can be crucial to better manipulate magnetic nanostructures for spintronics, as well as for medical applications.

**Author Contributions:** Conceptualization K.C.; Nanoparticle fabrication R.H., M.T., D.G., and B.N.; TEM imaging and VSM magnetometry K.C., Y.C., D.G., M.T., and B.N.; synchrotron XMCD and XRMS measurements K.C., D.G., D.S., Y.C., T.W., T.L, E.J., and A.H.R.; XRMS data reduction D.G. and Y.C.; XMCD and XRMS data analysis and modeling K.C., D.G., D.S., B.N., and S.K.; Original draft preparation, K.C.; Reviewing & editing K.C., R.H., A.H.R., and all co-authors.

**Funding:** Use of the Stanford Synchrotron Radiation Light Source, at SLAC National Accelerator Laboratory, was supported by the U.S. Department of Energy, Office of Science, Office of Science User Facility. T.W., T.L., E.J., and A.H.R. were supported by the Department of Energy, Office of Science, Basic Energy Sciences, Materials Sciences and Engineering Division, under Contract DE-AC02-76SF00515.

**Acknowledgments:** This research, including the sample preparation, TEM and VSM measurements, as well as synchrotron experiments, was funded through BYU MEG grants. Sample fabrication, and TEM and VSM measurements were carried out at BYU facilities. We thank Jeffrey Farrer and Paul Minson for their help with the collection of the TEM images. We are deeply grateful to Jeffrey Kortright for helpful discussions and for sharing his precious expertise in magnetic scattering.

**Conflicts of Interest:** The authors declare no conflict of interest

## Appendix A

The extraction of magnetic information from the XRMS signal via the dichroic ratio $R_D$ and magnetic ratio $R_M$ is based on the theory of magnetic scattering [39,40,52]. In this theory, the scattering factor *f* may be written as the sum of two terms:

$$f_\pm = f_c \pm f_m \tag{A1}$$

where $f_c$ represents charge scattering, $f_m$ represents magnetic scattering, and the ± sign refers to the helicity of the circularly polarized light (clockwise and counter-clockwise, respectively) in the plane transverse to the direction of propagation.

The scattering amplitude *A* produced by the ensemble of atoms in the nanoparticle assemblies at a specific direction *q* in the scattering space is then expressed as:

$$A_{\pm}(q) = \sum_j (f_{c,j} \pm f_{m,j})e^{i\vec{q}\cdot\vec{r}_j} = f_c s_c \pm f_m s_m = A_c \pm A_m \quad (A2)$$

where the sign $\sum$ refers to a summation over the atoms, indexed by *j*. The expression for *A* is here simplified by introducing a charge and a magnetic structure factors $s_c$ and $s_m$, both complex numbers, which respectively depend on the spatial distribution of the charge and the magnetism in the matter, and which are both *q*-dependent. Altogether, the scattering amplitude *A* can simply be expressed as a sum between a charge amplitude $A_c$ and a magnetic amplitude $A_m$, all of which are complex numbers. The scattering intensity *I*, a real quantity that is eventually measured on the detector, is then expressed as following:

$$I_{\pm} = |A_{\pm}|^2 = |A_c|^2 \pm (A_c A_m^* + A_m A_c^*) + |A_m|^2 \quad (A3)$$

In the scattering intensity, one can identify a pure charge term $|A_c|^2$, a pure magnetic term $|A_m|^2$, and a cross term, mixing charge magnetism, whose sign depends on the polarization of the light. Information about the magnetic distribution can then be extracted by exploiting the dependence with light polarization. Scattering patterns are collected in opposite light helicities and their respective intensities *I+* and *I-* are compared. The dichroic ratio is computed as following:

$$R_D = \frac{I_+ - I_-}{I_+ + I_-} \quad (A4)$$

Given that the magnetic component $A_m$ is usually small in respect to $A_c$, it can be shown that the dichroic ratio is related to both $s_c$ and $s_m$, as following (after dropping a phase shift):

$$R_D \approx \frac{2Re(A_c A_m^*)}{|A_c|^2} \propto \left|\frac{s_m}{s_c}\right| \quad (A5)$$

This dimensionless dichroic ratio $R_D$ allows one to quantify the relative strength of the magnetic contribution with respect to the charge contribution. However, the implication of both structural factors $s_c$ and $s_m$ make the extraction of the magnetic information complicated.

Another ratio, which we call here the "magnetic ratio" $R_M$ is defined this way:

$$R_M = \frac{I_+ - I_-}{\sqrt{(I_+ + I_-)}} \quad (A6)$$

Under the same approximations, one can show that $R_M$ is directly proportional to the magnitude of the complex magnetic factor $s_m$ (after dropping a phase shift):

$$R_M \approx \frac{2Re(A_c A_m^*)}{|A_c|} \propto |s_m| \quad (A7)$$

Since the scattering intensities $I_+$ and $I_-$ are measured at different *q* values, the ratio $R_M$ is also a function of *q*. Plotting $R_M(q)$ gives a profile directly proportional to the magnetic structure factor $s_m(q)$, thus revealing the pure magnetic ordering.